\documentstyle[twocolumn,prl,aps]{revtex}
\begin{document}
\draft
\title{Fractional Exclusion Statistics and Two Dimensional Electron Systems }
\author{ R.K.Bhaduri, M. V. N. Murthy\thanks{Permanant Address: The
Institute of Mathematical Sciences, Madras 600 113, India}
and M.K.Srivastava\thanks{Department of Physics, University of Roorkee,
Roorkee, India}}
\address
{Department of Physics and Astronomy, McMaster University, Hamilton,
Ontario, Canada L8S 4M1}
\date{\today}
\maketitle
\begin{abstract}
Using the Thomas-Fermi approximation, we show that an interacting two
dimensional electron gas may be described in terms of fractional
exclusion statistics at zero and finite temperatures when the
interaction has a short-range component. We argue that a likely
physical situation for this phenomenon to occur may exist in two
dimensional quantum dots.
\end{abstract}
\narrowtext

Fractional exclusion statistics or the generalised exclusion principle was
first proposed by Haldane\cite{H91a,H91b} in the context of excitations in
spin chains. Experimentally, the best evidence comes from recent neutron
inelastic scattering experiment\cite{CY95} on the compound $KCuF_3$, which
is a one-dimensional Heisenberg antiferromagnet above $40^0 K$. The
observed inelastic scattering is best fitted by spinon excitations in a
spin chain whose pairwise interaction falls off as the inverse square of
the lattice distance\cite{MU81}. The dynamic correlation function for such
a system has been calculated by Haldane and Zirnbauer\cite{HZ93}.  The
concept of fractional exclusion statistics has been generalised to the
case of a gas of particles\cite{MS94} defined by a distribution function
\cite{W94,I94} that allows for partial or multiple occupancy of a
single-particle state. In principle, the statistics is applicable to
particles in any spatial dimension, but most known examples are
mathematical models in one dimension \cite{C69,S71,SS71} with pairwise
inverse-square interaction. The first calculation for a two-dimensional
realistic system in this context was done by Johnson and
Canright\cite{JC94}, who demonstrated, by exact diagonalisation of a small
number of interacting electrons, that the bulk excitations in FQHE liquids
exhibit Haldane statistics. In this paper, we show that under certain
conditions, a two-dimensional interacting electron gas in its ground state
may exhibit this statistics. The conditions are shown to be favourable for
electrons in a quantum dot. In this case, it is shown that the dominant
effect of the interaction may be incorporated in the fractional statistics
of the gas. If the residual interactions are neglected, then that the
system also obeys Haldane statistics at finite temperature. This opens up
the exciting possibility that the bulk properties of a mesoscopic
two-dimensional system may be understood by regarding it as an almost
ideal fractional statistics gas confined in a potential well.

The claims made in this paper are based on the Thomas-Fermi (TF)
method\cite{T28}. Being a mean-field method, it cannot reproduce
two-body correlations, but is successful in giving a good estimate of bulk
properties like the ground-state energy and the single-particle
spatial density. It
has previously been applied with success to atoms\cite{M75},
nuclei\cite{B85}, and metal clusters\cite{B93}.  In two-dimensions, TF
yields an accurate approximation to the total energy of a many-anyon
system\cite{LBM92}. For an ideal gas obeying the generalised exclusion
statistics, TF calculation has been shown to yield the exact answer for
the energy in the large-N limit\cite{SB95}.  It is therefore reasonable to
expect that the method gives meaningful answers.  We start by constructing
the energy density functional for the ground state energy of a system of
interacting spin-half fermions.  Consider the N-fermion Hamiltonian in two
dimensions:
\begin{equation} H~=~{1\over{2m^*}}\sum_{i=1}^{N} p_i^2
+\sum_{i=1}^{N}V_1(r_i) + \sum_{j<k} V_2(|\vec r_j-\vec r_k|),
\end{equation}
where $V_1$ is a one-body confining potential whose
specific form is not crucial at present and $V_2$ is the two body
potential which is  repulsive. In a mean-field theory, the
expression for the energy at zero-temperature is given by,
\begin{eqnarray}
E~&=&~\int d^2 r\left [\frac{\hbar^2}{2m^*} \tau(r) \right.
+V_1(r)\rho(r) \nonumber \\
& & + {1\over 2}\{\rho(r)\int d^2r' \rho(r') V_2(|\vec
r-\vec  r~'|)~ \nonumber \\
& & \left.  - C\;\int d^2 r'|\rho(r,r')|^2 V_2(|\vec r-\vec r~'|)\} \right],
\end{eqnarray}
where $\rho(r)$ is the spatial single-particle density, $\tau(r)$ is the
kinetic energy density and $\rho(r,r')$ is the density matrix. In the
above we have taken into account the effect of both direct and exchange
terms in the interaction energy.  The factor 1/2 is the correction due to
over-counting of pairs. The constant $C$ is determined by the spin
polarisation of the gas : for unpolarised electrons, it is $1/2$, whereas
for a fully polarised system, it is $1$. For arbitrary polarization
$P=\frac{N_+ -N_-}{N}$, where $N_{\pm}$ is the number of up or down
spins,  the factor $C=\frac{1+|P|}{2}$. The spatial density is normalized
such that $N = \int d^2 r \rho(r)$. In the Thomas-Fermi method, the
kinetic energy density $\tau(r)$ is itself expressed in terms of the
density $\rho(r)$ and its gradients. The energy and the density are
determined self-consistently by a variational principle. In two-dimensions,
the TF expression for $\tau(r)=\pi\rho^2 (r)$, taking into account the
spin-degeneracy factor of 2. In this case, there is no gradient
correction in the bulk up to $O(\hbar^2)$. However, there are edge
corrections when the sample is of finite size\cite{JB75}.

Next consider the energy due to the two body interactions.
The matrix elements of the direct term is,
\begin{equation} \sum_{i,j} <ij|V_2|ij> =\int
\rho(r_1)\rho(r_2) V(|\vec r_1-\vec r_2|) d^2r_1 d^2r_2,
\end{equation}
where the sum (here as well as in what follows) is over the {\it occupied
single-particle states} only.
The matrix elements of the exchange term is
\begin{equation}
\sum_{i,j} <ij|V_2|ji> =\int
|\rho(r_1,r_2)|^2 V(|\vec r_1-\vec r_2|) d^2r_1 d^2r_2,
\end{equation}
where $\rho(r_1,r_2)=\sum_i\psi^*_i(r_1) \psi_i(r_2)$. At this stage, it
is useful to perform
the density-matrix expansion following Skyrme\cite{S56}.
Defining $\vec r = \vec r_1 -\vec r_2$
and $\vec R = (\vec r_1 +\vec r_2)/2 $ and expanding the density up to
this order in $\vec r$ , we obtain
\begin{equation}
\rho(\vec r_1) = \rho(\vec R + \vec r/2) = \rho(\vec R)  +(\vec
r.\nabla)\rho +{1\over 2}(\vec r.\nabla)^2 \rho+\ldots
\end{equation}
The direct matrix element may then be written as,
\begin{eqnarray}
\sum_{i,j}<ij|V_2|ij>~=~ \int d^2 r V_2(r) \int d^2 R \rho^2(R)
\nonumber \\
- {1 \over 4} \int d^2 r~ r^2~ V_2(r) \int d^2 R (\nabla \rho(R))^2+\ldots,
\end{eqnarray}
Similarly  the density matrix $\rho(\vec r_1,\vec r_2)$ may be expanded up
to second order in $\vec r$ about $\vec R$,
\begin{eqnarray}
& &\rho(\vec r_1, \vec r_2) = \sum_i \psi^*_i(\vec R+\vec r/2) \psi_i(\vec R
-\vec r/2)\nonumber \\
&=&\sum_i \left [ \psi^*_i(\vec R)\psi_i(\vec R)\right.
\nonumber \\
&+&\left. {1\over 16} r^2(\psi_i^*\nabla^2 \psi_i
+(\nabla^2\psi^*_i)\psi_i  -2\nabla\psi^*_i.\nabla\psi_i)\right],
\end{eqnarray}
and  the exchange contribution to second order is given by,
\begin{eqnarray}
\sum_{i,j} <ij|V_2|ji> = \int d^2 r V_2(r) \int d^2 R \rho^2(R)
\nonumber \\
- {1 \over 2} \int d^2 r~ r^2~ V_2(r) \int d^2 R \tau(R) \rho(R) +\ldots.
\end{eqnarray}
Here the kinetic energy density is defined as,
\begin{equation}
\tau = -{1\over 4} \sum_i( \psi_i^*(\nabla^2 \psi_i)
+(\nabla^2\psi^*_i)\psi_i) +{1\over 2}\sum_i(\nabla \psi_i^*).(\nabla
\psi_i).
\end{equation}
Often the kinetic energy density is defined either by the first term or
by the second term in the above equation without the
over all 1/2.  What we naturally get in the expansion
is an average of both these commonly used forms. We have
computed each one of these forms exactly using harmonic oscillator wave
functions for a few particles. While the the first and second terms show
oscillations around the smooth TF density, the definition given above
almost precisely coincides with the TF density even with as little as two
particles.

We note that the leading terms in both direct and exchange terms
are the same (proportional to $\rho^2$). For
spin-half fermions the interaction energy is given by,
\begin{equation}
\sum_{i,j} [<ij|V_2|ij> - \delta_{m_i,m_j}\delta_{m_j,m_i}<ij|V_2|ji>],
\end{equation}
where $m_i$ is the spin projection.  Summing over all particle indices
immediately gives a factor $(1+|P|)/2$ for the exchange contribution,
where $P$ is the spin polarization of the system.
Therefore, if there is no other degree of freedom, or if the spins are
all polarized, the contribution from the leading terms to the
interaction energy vanishes as it happens in FQHE
systems.
However for the unpolarized 2-D electron systems  there is a factor
half for exchange contribution. Here we concentrate on the
unpolarized case.  Combining all the contributions the total energy of
the system is given by,
\begin{eqnarray}
E~=~\int d^2 r\left [\frac{\hbar^2}{2m^*} \pi\rho^2(r)
+V_1(r)\rho(r) + {1\over 4} \rho^2(r) M_0 \right. \nonumber\\
\left. +{1\over 8}[(\pi\rho^3(r) -(\nabla \rho(r))^2] M_2 +\ldots
\right],
\end{eqnarray}
where $M_n=\int d^2r V_2(r) r^n$ are the moments of the two body
potential. Note that we obtain an expression similar to the above if we
use an expansion of the form\cite{TK85},
\begin{equation}
V_2(r) = \sum_{j=0} c_j b^{2j} \nabla^{2j} \delta^2(\vec r),\label{eff}
\end{equation}
where $b$ is the range of the potential and $c_j$ are related to the j-th
moment of the potential $V_2$ as $ M_{2j} = 2^{2j}j!c_j b^{2j}$.

The spatial density is now determined by the variation $\delta (E-\mu
N)=0$, where $\mu$ is the chemical potential at zero temperature.  The
variation immediately gives the equation for the density,
\begin{eqnarray}
& &\frac{\pi\hbar^2}{m^*}\left [1 + \frac{m^*M_0}{2\pi\hbar^2}\right]
\rho(r) +\frac{3\pi M_2}{8}\rho^2(r) +\frac{M_2}{4}\nabla^2 \rho(r)
\nonumber \\
&=& \mu-V_1(r).
\label{eqnrho}
\end{eqnarray}
In the large-N  limit we expect the density in the bulk to be approximately
constant. We can therefore neglect the derivative term in this limit.
Further if the potential is extremely short-ranged, the term proportional
to the second moment of the potential may also be neglected.
(We will elaborate on these approximations shortly.)
Then the density is given by,
\begin{eqnarray}
\rho_0(r) &=& \frac{m}{\pi\hbar^2\alpha} (\mu -V_1(r)),~~r\le r_0 \nonumber\\
          &=& 0,~~~~~~~~~~~~~~ r> r_0, \label{rhonot}
\end{eqnarray}
where $r_0$ is the classical turning point defined by $\mu=V_1(r_0)$ and
\begin{equation}
\alpha = 1+\frac{m^*M_0}{2\pi\hbar^2} \label{stat}
\end{equation}
is now the statistics parameter as we show below.
In the effective range expansion (\ref{eff}), $c_0=M_0$.
The expression for $\rho_0$ in Eq.(\ref{rhonot}) may be interpreted
as if the fermions in the one-body confining potential $V_1$ are
noninteracting, but that they{\it obey the generalised exclusion statistics
for occupancy} at zero temperature:
\begin{eqnarray}
n(\epsilon) &=&{1\over \alpha},~~~ \epsilon < \mu \nonumber \\
	    &=& 0,             ~~~ \epsilon > \mu. \label{occ}
\end{eqnarray}
This may be easily seen as follows. For noninteracting fermions,
the Thomas-Fermi density of states $g(\epsilon)$ in an external
potential $V_1(r)$ is
\begin{equation}
g(\epsilon)= 2 \int \frac{d^2r d^2p}{(2 \pi \hbar)^2}\,\, \delta \left(
\epsilon-\frac{p^2}{2m^*}-V_1(r)\right)\,.
\end{equation}
The over-all factor of two on the right-hand side is due to the spin
degeneracy. Using the new occupancies given by Eq.(\ref{occ}), we
get
\begin{eqnarray}
N&=&\frac{1}{\alpha}\int_{0}^{\mu} g(\epsilon)d \epsilon \,\nonumber \\
 &=&\frac{1}{\alpha} \int 2 \; \frac{d^2r d^2p}{(2 \pi \hbar)^2} ~~\theta
\left(\mu-\frac{p^2}{2m^*}-V_1(r)\right)\,. \end{eqnarray} The function
$\theta(y)=1$ for $y>0$, and zero otherwise. Now performing the
p-integration immediately yields the total number of particles, with
density $\rho_0(r)$ given by Eq.(\ref{rhonot}). Indeed we have now the
precise condition under which ideal exclusion statistics is realised
within the framework of the Thomas-Fermi method.

In the more realistic situation, the higher moments may not be neglected,
and the system is a non-ideal fractional statistics gas. In the thermodynamic
limit, we may write
\begin{equation}
\rho(r) = \rho_0(r)\left[ 1 -\frac{3m^*M_2}{8\hbar^2\alpha}\rho_0(r)
+\ldots \right],
\end{equation}
where $\rho_0(r)$, given by Eq.(\ref{rhonot}), is the density for the
ideal FES case. Note that $M_2 = 4c_1 b^2$ where $b$ is the range of the
potential. The typical densities in two dimensional systems of interest
is
of the order of $10^{-5}/\AA^2$.  Using the values of $m^*=0.067m_e$,
which is the effective electron mass in GaAs materials, and $\alpha \ge 1$
(but not very large), it is easy to estimate that the second term becomes
important only for ranges of the order of $100\AA$ or above.
Another way to view the problem is to regard the short-range part of the
two-body interaction, which dominates $M_0$, to alter the statistics only.
The long-range part of $V_2$, giving the higher moments, modifies the
self-consistent mean field. Consider for example the electrons in two
dimensional quantum dots. The two body potential is usually taken to be
the Coulomb interaction and the confining potential of the device is
modelled by the oscillator potential. However, it is expected that the
effective two-body interaction after averaging over the probability
densities in the direction perpendicular to the plane will be more
complicated. Many qualitative features of the system may be explained by
several choices of the potential. As in the case of FQHE liquids, we
assume that the model interaction has a short range part $V_{2s}(r)$ and a
long range part $V_{2l}(r)$. We use the moments expansion for the
short-range part and neglect the effect of higher moments. The
self-consistent equation for the density is then given by,
\begin{eqnarray}
\rho(r) &=& \frac{m^*}{\pi\hbar^2\alpha} (\mu -U(r)),~~r\le r_0 \nonumber\\
          &=& 0,~~~~~~~~~~~~~~~~~~ r> r_0, \label{rho}
\end{eqnarray}
where the mean TF potential is defined as
\begin{equation}
U(r) = V_1(r) + \int d^2 r \rho(r') V_{2l}(|\vec r - \vec r~'|).
\end{equation}
The equation further simplifies for circularly symmetric density. Expanding
the potential in partial waves,
$$V_{2l}(|\vec r -\vec r~'|) = {1\over \pi}\sum_{m=0}^{\infty}
v_{m}(r,r')\cos m(\theta-\theta '), $$
the TF potential reduces to,
\begin{equation}
U(r) = V_1(r) + \int r' dr' \rho(r') v_0(r,r').\label{selfcon}
\end{equation}
In the  above equation we have ignored the exchange effects
which are not important for the long range potentials. Thus the
Eq.(\ref{selfcon}) is the self consistency condition to determine the
density $\rho(r)$, and in general is not solvable analytically.

Finally we  consider briefly the finite temperature problem using
the Thomas-Fermi method. We restrict our attention to the case where the
two body potential is extremely short-ranged and regard the system as
ideal.  The temperature $T$ is expressed in
units of the Boltzmann
constant, so that it has the dimensions of energy. The
one-body potential is now temperature-dependent, and is given by
\begin{eqnarray}
V(r,T)&=&V_1(r)+\frac{M_0}{2}\rho(r,T) \nonumber\\
&=& V_1(r)-(1-\alpha)\frac{\pi \hbar^2}{m^*}\,\rho(r,T)\,,\label{vtem}
\end{eqnarray}
where $\alpha$ is the statistics parameter defined by Eq.(\ref{stat}).
We have assumed that the external potential $V_1(r)$ is temperature
independent. In the above equation, the density $\rho(r,T)$ for the
fermions is obtained from the relation (including the spin-degeneracy of
2)
\begin{equation}
\rho(r,T)=\frac{2}{(2 \pi \hbar)^2}\int \frac{d^2p}
{\exp[(p^2/2m^*+V-\mu)/T]+1}\;,
\end{equation}
and the chemical potential is determined by $N=\int d^2r \rho(r,T)$.
The $p-$integration above may be done analytically, giving
\begin{equation}
\rho(r,T)= \frac{m^*T}{\pi \hbar^2}\,\ln\left(1+\exp
[-(V-\mu)/T]\right)\;.
\end{equation}
This is inverted to give
\begin{equation}
\frac{\mu}{T}=[V+\frac{\pi \hbar^2}{m^*}
\rho]/T\,+\,\ln\,\left(1-\exp(-\pi
\hbar^2 \rho/m^*T)\,\right)\;.
\end{equation}
Substituting for $V$ above from Eq.(\ref{vtem}), we get
\begin{eqnarray}
\frac{\mu}{T}&=&\left(V_1(r)+\alpha \frac{\pi \hbar^2}{m^*}
\rho(r,T)\right)/T \nonumber\\
&+&\ln\,\left(1-\exp (-\pi \hbar^2 \rho/ m^*T)\,\right)\,.
\end{eqnarray}
For a gas in the thermodynamic limit, we set $V_1(r)=0$ above. Further, the
spatial density $\rho$ may be expressed as $2 \rho_{0}$, where $\rho_{0}$
is the density for spin-less partcles. Then Eq.$(27)$ reduces to the form
\begin{equation}
\frac{\mu}{T}=\alpha \frac{2 \pi \hbar^2}{m^*T} \rho_{0} +
\ln\,\left(1-\exp (-2\pi \hbar^2 \rho_{0}/ m^*T)\,\right)\,.
\end{equation}
Note that this is precisely the equation derived by Wu\cite{W94}
(see his Eq.$(23)$ )
for a two-dimensional gas obeying the statistics
\begin{equation}
n(\epsilon)=\frac{1}{w(\exp (\epsilon-\mu)/T)+\alpha}\;,
\end{equation}
with $w(x)$ satisfying the functional equation
\begin{equation}
w^{\alpha}(1+w)^{1-\alpha}=x=\exp(\epsilon-\alpha)/T\;.
\end{equation}
Here, as in our case, $\alpha=1$ corresponds to free fermions.

We have thus shown that in the large-N limit, ideal exclusion statistics
may be realised in a system of spin-half fermions with very short-range
interactions.   Note that this situation is peculiar to two
dimensions since both the leading term in the moments expansion and the
kinetic energy density have the same dependence on the spatial density.

We thank Diptiman Sen for comments. This research was supported by a grant
from NSERC(Canada). M.V.N. and M.K.S thank the Department of
Physics and Astronomy, McMaster for Hospitality and R.K.B thanks
Hank Miller, University Pretoria where part of the work was done under
a grant from the Foundation for Research and Development of South Africa.

\end{document}